# Binding Energies of the α Particle and the A=3 Isobars from a Theoretical Geometric Model.

Gustavo R. González-Martín[a]

**Departamento de Física, Universidad Simón Bolívar, Apartado 89000, Caracas 1080-A, Venezuela.**

We assume a triple geometric structure for the electromagnetic nuclear interaction. This nuclear electromagnetism is used to calculate the binding energies of the alpha particle and the A=3 isobar nuclides. The approximation for the resultant wave equation which lead to the deuteron binding energy from the modified Mathieu equation for the radial eigenvalue equation also establishes proton-electron-proton magnetic bonds in these nuclides and determines their binding energies. Completely theoretical calculations give *28.5 Mev., 7.64 Mev.* and *8.42 Mev.* for the binding energies of the alpha particle $^4$He, the $^3$He and $^3$H nuclides respectively. These values admit correction factors due to the approximations made.

21.10Dr, 21.60.-n, 14.20.Dh, 02.20.Qs

[a] Webpage http://prof.usb.ve/ggonzalm

# 1- Introduction.

In a previous paper [1] we calculated the binding energies of the deuteron from a geometric theoretical model. A modified Pauli equation [2,3,4] with a non standard electromagnetic coupling introduces an SU(2) structure. This equation was also useful in understanding the proton and neutron magnetic moments [3]. Electromagnetism is represented by a connection potential [5] corresponding to the $SU(2)_Q$ generators which are subject to quantization in the same manner as the angular momentum generators. In particular, this geometric coupling provides an attractive $1/r^4$ magnetic potential that may be important in nuclear processes.

With the approximations made, the modified Pauli equation for a proton-electron-proton magnetic bond lead to the Mathieu equation [6,7,8]. It is known that the characteristic roots and the corresponding series expansion coefficients for the Mathieu functions have branch cut singularities [8,9,10]. To eliminate the possibility of multiple-valued expansion coefficients in the even Mathieu functions of period $\pi$ and obtain a regular solution, we disregarded all points $q$ on all branches of the Riemann surface except well known common points $q_0$ [10,11,12]. The binding energy was obtained in terms of this Mathieu parameter $q_0$.

The presence of orbital angular momentum, $\lambda \neq 0$, breaks the stable quasi-static mechanism of the model. The equation determines only one stable magnetic bond excitation corresponding to the states $\nu = \pm \frac{1}{2}$ of frame field excitations bound by the strong magnetic interaction.

This deuteron "*pep* magnetic bond" is a fundamental excitation and supplies a coupling mechanism which allows the combination of more than 2 protons. The strong magnetic fields of two different *pep* frame magnetic bonds may interact. Using this ideas the model was extended to the alpha particle or *4He* nuclide considered as a 2-deuteron excitation [1]. The extended *4He* model presented is not satisfactory because it requires additional unnecessary assumptions. In this article we continue the previous discussion and present a better $\alpha$ model which also applies to the A=3 isobar nuclides *3H* and *3He* using the same geometric ideas [1] in terms of the *pep* magnetic bond.

# 2- The Proton-Electron-Proton *pep* Magnetic bond.

We apply the modified Pauli equation to a system of geometric excitations, in particular to a system of one electron and two protons moving about the system center of mass. The fields are dominated by the electron magnetic field because its higher magnetic moment $\mu$. The resultant SU(2) field is characterized by the system reduced mass which is the electron mass $m$. The system $(p, e, \bar{\nu}, p)$ was used as a phenomenological a model for the deuteron as indicated in [1]. Nevertheless the fundamental magnetic nature of this system may be appreciated if we consider the magnetic flux linking the components. According to the geometric ideas there are flux quanta associated to the protons $p$ and the electron $e$, one for each particle, [13] that should link among themselves. This makes convenient to use the flux lines or magnetic strings to characterize the links of particle magnetic moments. The dominant potential is a strong attraction in the equatorial radial direction which may also be considered as the attraction among magnetic flux lines linked to flux quanta. The three indicated quantum flux strings cannot be linked so that they fully attract themselves as a whole without breaking the Pauli exclusion principle of the two $p$. To fully attract themselves the only possibility is the necessary presence of a fourth string. This string must be supplied by the only neutral fundamental excitation in the theory, an $L$ excitation or neutrino with a linked flux quantum. The resultant system $(p, e, \bar{\nu}, p)$ is the model for the deuteron. The only possible way to attractively link the flux lines determines that the subsystem $(e, \bar{\nu}) \equiv e'$ has the $e$ and $\bar{\nu}$ spins in the same direction. Hence $e'$ has spin 1, the charge, mass and the magnetic moment of the electron. The operators $^+A_e$ of the electron $e'$ and $A_p$ of the protons $p$ in the stable system $(p, e', p)$ constitute a single operator $A$, which depends on the reduced mass, and determines the fundamental SU(2) excitation. In other words, the resultant 4-potential $A$ has the Clifford algebra orientation required by the fundamental $SU(2)_Q$ quantum representation of the generator. This determines the geometric angle $\Theta$ of $A$ with respect to the even electromagnetic potential $^+A$. The resultant field affects the protons. The dominant internal magnetostatic energy is the one that holds the excitation system $(p, e', p)$ together and determines its energy proper value $E$ and the *pep* link binding energy which is [1]

$$U_d = -2q_0^2 m_e = 2.20474 \quad \text{Mev.} \tag{1}$$

# 3- The many Deuteron Model.

The strong magnetic interaction range [14] defines the subnuclear zone $r \sim 1/m_W \ll r_p$, very small in relation to the proton radius $r_p$. Mathieu's equation determines a single *pep* bond energy between the 2 protons of a pair. In order to feel the strong attraction the proton centers must be inside the subnuclear zone. Therefore the protons are essentially superposed. In the $\alpha$ particle model there are 4 protons and 2 electrons. The extra protons are also superposed and must share the magnetic field. The stationary state, of minimum energy, is the totally symmetrized quantum superposition, as in the covalent bond for the hydrogen molecular ion. The electrons and protons are shared by all possible *pep* magnetic bonds in a system as a quantum superposition. The possible *pep* magnetic bonds obey the Pauli exclusion principle in the $(p,e,\bar{\nu},p)$ system and therefore are only those sharing a single electron $e'$. All $p$ participate in as many *pep* magnetic bonds as possible proton pairings.

One difference between the alpha and deuteron models is that in the alpha model the reduced mass $m$ of the system is approximately half the electron mass $m_e$ due to the presence of two electrons in the system. **The required coordinate transformation to obtain Mathieu's equation** is then different from the deuteron model [1],

$$\rho^2 = \frac{\mu}{m\varepsilon}z^2 = \frac{\gamma}{m_e m\varepsilon}z^2 = \frac{\gamma}{2m^2\varepsilon}z^2 \,, \tag{2}$$

giving

$$R'' + \left[\gamma\varepsilon\left(e^{2u}+e^{-2u}\right)-\left(\nu^2-\alpha^2\right)\right]R = 0 \,, \tag{3}$$

which changes the parameter $q$ proportionally [1] and determines twice the binding energy $U_d'$ per pair,

$$q = \varepsilon/2 \,, \tag{4}$$

$$E_d' = 2E_0' = 2\varepsilon^2 m = 2(2q)^2 m = -4q_0^2 m_e = -2U_d \,. \tag{5}$$

The energy $E_d' = 2E_d$ is the unique Mathieu equation eigenvalue even if $e'$ is shared and is the same for each *pep* magnetic bond. This fact together with the symmetry simplifies the energy calculation of the *n*-body problem. There are symmetric cluster states whose energies are determined by their *pep* bonds. The symmetric $\alpha$ cluster associated to 4 protons has 6 *pep* bonds and a total Mathieu magnetic bond energy $E_\alpha$,

$$\sum_{p,e'}^{4p,2e'} m_i = \sum_{p,e'}^{4p,6e'} m_i - 4m_{e'} \,, \tag{6}$$

$$E_{(4p,2e')} = E_{(4p,6e')} - 4m_e = 6E_d' - 4m_e = -28.5009 \text{ Mev.} \equiv E_\alpha \approx -28.28 = -U_\alpha. \tag{7}$$

Similarly we have the same considerations for 3 protons. The symmetric cluster associated to 3 protons has 3 *pep* bonds with a unique Mathieu energy $E_d$,

$$E_{(3p,e')} = E_{(3p,3e')} - 2m_e = 3E_d - 2m_e = -7.6362 \text{ Mev.} \equiv E_{^3He} \approx -7.72 = -U_{^3He} \,. \tag{8}$$

The exclusion principle prohibits $2e$ in a *pep* magnetic bond. If there are $2e$ in the $3p$ system a neutron is created and

$$\Delta E = m_e - (m_n - m_p) \,, \tag{9}$$

$$E_{(3p,2e)} = E_{^3He} + \Delta E = -8.4163 \text{ Mev.} \equiv E_{^3H} \approx -8.48 = -U_{^3H} \,. \tag{10}$$

It should be noted that these relations determine a mass relation identity for the A=3 isobars. Since our notation uses the correspondence $(3p,e') = (2p,n)$ we may write

$$M_{^3H^+} = m_p + 2m_n + E_{^3H} = m_p + 2m_n + E_{^3He} + \Delta E = 2m_p + m_n + m_e + E_{^3He} \equiv M_{^3He^{2+}} + m_e,$$
**(11)**

which indicates a small $\beta$ decay possibility of $^3H^+$ into $^3He^+$ from total energy considerations in spite of the unfavorable binding energy difference. It also implies that the corresponding atoms $^3He$ and $^3H$ have essentially equal masses, up to the negligible Coulomb electron interaction.

We have obtained the binding energies for $^4He$ and the isobar nuclides $^3H$, $^3He$. Therefore, we are lead by the model to consider that the *pep* magnetic bonds or deuterons act as the essential components of more complex clusters in electromagnetic interaction with additional nucleons. This is consistent with the protonic and neutronic numbers of the nuclides. Either proton in a *pep* magnetic bond when associated with $e'$ may be considered a neutron. Therefore the energy bond $U_d$ determines a "nuclear" force among nucleons which is independent of a neutron association.

## 4- Conclusion.

The numbers obtained for the binding energies are surprisingly close to the experimental values, for this crude model. These values admit correction factors due to the approximations made. The *pep* bond plays a role similar to the *e* covalent bond in atomic physics.